\documentclass[12pt]{iopart}
\usepackage{graphicx}
\usepackage{amssymb}
\begin{document}

\title[Gapless excitations of vortices with axial symmetry]{Gapless excitations of axially symmetric vortices in systems with tensorial order parameter}
\author{A J Peterson$^{1}$ and M Shifman$^{1,2}$}

\address{$^{1}$ School of Physics and Astronomy, University of Minnesota, Minneapolis, MN 55455, USA \\
$^{2}$ William I. Fine Theoretical Physics Institute, 
University of Minnesota, Minneapolis, MN 55455, USA \\}
\ead{pete5997@umn.edu, shifman@umn.edu}

\begin{abstract}
We extend the results of previous work on the vortex order parameter in systems similar to the Ginzburg-Landau description of superfluid $^3$He in the bulk B phase.  Specifically, we consider vortices preserving an axial $U(1)$ symmetry.  We determine the conditions required by the $\beta_i$ parameters to allow for an energetically favorable development of the off-diagonal antisymmetric and symmetric-traceless elements satisfying the axial symmetry from the trace-only ansatz of the order parameter.  The number and type of gapless moduli appearing on the classical low energy theory of axial vortices is determined.  The time-dependent part of the Ginzburg-Landau free energy is then considered to determine the number of quantized modes emerging from the gapless modulus fields.
\end{abstract}
\submitto{\JPCM}
\maketitle

\section{Introduction}
In a previous publication \cite{Peterson:2013zba} we considered emergent modulus fields on mass vortices in systems similar to the Ginzburg-Landau description of superfluid $^3$He-B.  In that work we provided a method for determining the type and number of moduli appearing on mass vortices under the restriction that the $3 \times 3$ matrix order parameter contained only its diagonal $(\delta_{\mu i})$ and antisymmetric  off-diagonal $(\varepsilon_{\mu i k}\chi_k)$ elements, setting its symmetric-traceless part to zero.  Although this restriction was not necessarily experimentally sound, it allowed the calculations to be carried out with relative ease while still providing enough complexity to illustrate the development of non-Abelian moduli and their interactions with the already well studied translational moduli (Kelvin modes) \cite{Thomson:1880, Sonin:1987zz, Simula:2008a, Fonda, Krusius:1984a, Pekola:1985a, Hakonen:1983a, Kobayashi:2013gba}.  The presentation of that work was also purely classical and we made no attempt to discuss time dependence or quantization, which may alter the number of gapless excitations determined from the modulus fields.  In this work we aim to continue that analysis and discuss vortex solutions that are more closely related to certain types of vortices studied both experimentally \cite{Krusius:1984a, Pekola:1985a, Hakonen:1983a} and theoretically \cite{Thuneberg:1986a, Volovik:1986a} in superfluid $^3$He-B.  Specifically we will discuss vortices respecting an axial $U(1)$ symmetry, which we will define below.

Superfluid $^3$He has drawn much attention from the high-energy physics community due to its non-Abelian group structure and the tensorial nature of its order parameter \cite{Volovik:2006a,Babaev:2001zy}.  The topological excitations derived from the group structure share many close similarities to the non-perturbative solutions from Yang-Mills theories.  Indeed the analysis of the symmetry breaking in superfluid $^3$He results in a low energy field theory describing the dynamics of gapless excitations of the mass vortices that is similar in form to the low energy description of flux-tubes presenting Abrikosov-Nielsen-Olesen (ANO) \cite{Abrikosov:1957, Nielsen:1973} string-like solitons in Yang-Mills theories \cite{Gorsky:2004ad, Hanany:2003hp, Auzzi:2003fs, Shifman:2004dr, Hanany:2004ea,Eto:2005yh}.  In particular, when the phenomenological $\beta_i$, $\alpha$, and $\gamma_i$ parameters satisfy certain constraints (e.g. $\gamma_{2,3} = 0$) the moduli excitations follow a variation of the ANO string excitations.  The low energy sector of the vortex excitations is composed of a translational modulus field part resulting in Kelvin excitations, as well as an internal non-Abelian part, which typically appears in the form of an O(3) sigma model \cite{Shifman:2012zz,Shifman:2013oia,Nitta:2013mj} .  Additionally, when $\gamma_{2,3}$ are small the pattern by which certain moduli develop mass gaps is similar to the case of the ANO string where Lorentz symmetry breaking terms in the Lagrangian generate mass gaps for the non-Abelian modes \cite{Monin:2013kza}.  This example is of course illustrating the universality of low energy effective field theories in condensed matter, high energy, and cosmological systems \cite{Volovik:2006a}.

We hasten to mention that the requirements of $\gamma_{2,3} \rightarrow 0$ are different from the values in superfluid $^3$He approximated from the weak coupling BCS theory as $\gamma_1 = \gamma_2 = \gamma_3$ with strong coupling corrections that have been calculated in \cite{Choi:2007a}.  In this sense we are discussing only systems similar to the Ginzburg-Landau description of superfluid $^3$He, however we are not strictly adhering to the descriptions given from condensed matter theory and experiment.  Thus the results we obtain should be taken only as illustration, and should not necessarily be considered for experimental value.  We however point out that the prescription of $\gamma_{2,3} = 0$ can be approximately achieved in an ultra-cold fermi gas with p-wave pairing.  This of course is not the case for superfluid $^3$He. 

The symmetry structure of superfluid $^3$He can be determined by considering the microscopic BCS theory of the helium atoms at the critical temperature.  Below this critical temperature the individual helium atoms condense into Cooper pairs similar to the BCS description of superconductivity \cite{Leggett:1972a}.  However, for the case of $^3$He the short range hard core potential requires the $^3$He atoms to pair in an orbital p-wave.  To preserve the antisymmetric pairing requirement for fermions the Cooper pairs form a spin triplet state.  Thus the order parameter describing the superfluid states is given by a complex $3 \times 3$ matrix $e_{\mu i}$, where $\mu$ and $i$ describe the spin and orbital degrees of freedom respectively \cite{Leggett:1975a,Thuneberg:1987a,Sauls:1981}.  The continuous group describing the internal symmetry structure of superfluid $^3$He is thus given by
\begin{equation}
G = U(1)_P\times SO(3)_S \times SO(3)_{L}
\end{equation}
where the $U(1)_P$ represents the phase symmetry of the order parameter, and the groups $SO(3)_S$ and $SO(3)_{L}$ represents the symmetries of the spin and orbital degrees of freedom respectively.  This is in addition to the translational symmetries as well as the discrete time reversal and parity symmetries.  As we will see below it will be necessary to distinguish between internal orbital rotations and external coordinate rotations (see the example in \cite{Bedaque:2003}).  By external coordinate rotation, we are referring to a transformation resulting from a rotation of the coordinate system without a corresponding rotation to the orbital index.  We will consider the orbital index as an internal degree of freedom.  This distinction is important for the case of $\gamma_{2,3} = 0$ since in this limit the free energy gradient terms receive an enhanced symmetry $SO(3)_L \rightarrow SO(3)_{L_{\rm ext}} \times SO(3)_{L_{\rm int}}$.  This situation is reminiscent of the theory of elasticity where an unphysical vanishing of the bulk modulus leads to an enhanced symmetry of rotations $O(2) \rightarrow O(2) \times O(2)$ leading to the equivalence of scale and conformal transformations \cite{Riva:2005gd}.  In this work it will be necessary to consider the entire symmetry group including both the internal and external symmetries (see Figures 1 and 2).

\begin{figure}[h!]
\centering
\includegraphics[width=0.7\linewidth]{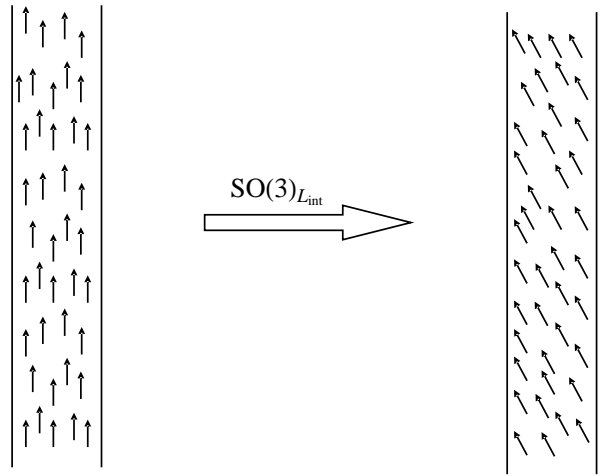}
\caption{The result of an internal rotation of the vortex orbital index is shown.  The vortex density function does not change, however the directors of the vortex are rotated.}
\end{figure}

\begin{figure}[h!]
\centering
\includegraphics[width=0.7\linewidth]{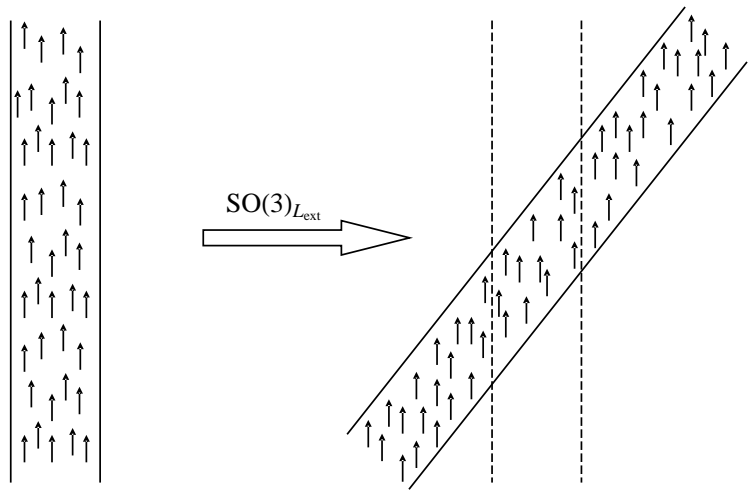}
\caption{The result of an external coordinate rotation of the vortex axis is shown.  The directors of the vortex solution however are not rotated.  A complete $SO(3)_L$ rotation would rotate both the density function and the directors.}
\end{figure}

There are several potential phases of the superfluid ground state, which are determined by considerations of the broken symmetries.  In particular the B phase is characterized by the spontaneous breaking of $G$ to a spin-orbit locked phase
\begin{equation}
G \rightarrow H_B = SO(3)_{S+L}
\end{equation}
similar to the mechanism of color-flavour locking in color superconductivity \cite{Alford:2007xm, Eto:2013hoa}.  The degeneracy of the B phase ground state is given by
\begin{equation}
G/H_B = U(1)_P \times SO(3)_{S-L}.
\end{equation}
This type of degeneracy allows for the existence of topologically stable vortices of the $\mathbb{Z} \times \mathbb{Z}_2$ type \cite{Lepora:1999}.  The specific vortex solutions are determined by minimizing the Ginzburg-Landau free energy for the required boundary conditions.

To accomplish this task we consider several forms of the order parameter and minimize the free energy under those assumptions.  Previously the vortex order parameters were searched for by initially decomposing the order parameter into its trace, symmetric, and antisymmetric components \cite{Peterson:2013zba,Nitta:2013mj}
\begin{eqnarray}
e_{\mu i} &=\frac{1}{3}e_{\sigma \sigma} \, \delta_{\mu i} + e^S_{\mu i} + e^A_{\mu i}, \nonumber \\
e^S_{\mu i} &= e_{\{\mu i\} }-\frac{1}{3}e_{\sigma \sigma}\delta_{\mu i}, \nonumber \\
e^A_{\mu i} &= e_{[\mu i]} \equiv \varepsilon_{\mu i k}\chi^k.
\label{CompleteDecomposition}
\end{eqnarray}
For the illustrative purposes in the previous work \cite{Peterson:2013zba} it was sufficient to consider just the trace and antisymmetric components of the order parameter
\begin{equation}
e_{\mu i} = {\rm e}^{{\rm i} \phi}f(\vec{x}_\perp)\delta_{\mu i}+\varepsilon_{\mu i k}\chi_k(\vec{x}_\perp),
\end{equation}
where $\vec{x}_\perp \equiv (x,y)$ are the coordinates in the plane perpendicular to the vortex axis, and $\phi$ is the polar angle in this plane.  The functions $f$ and $\chi$ are determined by minimizing the free energy.  This is indeed the most simple context allowing for the maximal number of modulus fields to emerge on the vortex.  In the present work we wish to continue that analysis by extending the order parameter ansatz to include all the necessary components appearing in an axially symmetric solution.  These will include new symmetric-traceless components.  In the absence of these terms the low energy effective theory can be described by translations (Kelvin modes) and rotations of the $\chi^i$ fields resulting in an emergent $O(3)$ sigma model for the unit vector $\lambda^i$ describing the director of $\chi^i$.  The symmetric terms will however complicate this situation since the symmetric components must be built from symmetric traceless combinations of the products $\lambda^\mu \lambda^i$ whereas the antisymmetric components are simply built from $\varepsilon_{\mu i k}\lambda^k$.  We will find that this addition does not change the number of moduli appearing on the mass vortex.  However, we will observe changes to the gradient and interaction terms involved.

The other goal of this paper will be to characterize the time dependence of the modulus fields to determine the quantized modes appearing on the mass vortex.  It is well known that after quantization the two translational moduli appearing on the vortex solution result in a single Kelvin mode.  This is the result of the application of Goldstone's theorem \cite{Goldstone:1961, Goldstone:1962} to non-relativistic systems where the number of quantized modes is less than or equal to the number of broken group generators \cite{Nielsen:1976, Watanabe:2012, Hidaka:2012ym}\footnote{Strictly speaking for a non-homogeneous vacuum degeneracy space the number of gapless modes may exceed the number of broken generators if flat directions corresponding to hidden symmetries are present  \cite{Novikov:1982}.  In most cases these modes become gapped when quantum corrections are included.  However, for some weakly coupled systems the corrections may be neglected and the modes associated with the hidden symmetries may be considered nearly gapless.  This effect occurs in the weak coupling limit of superfluid $^3$He-A \cite{Volovik:1982, Volovik:1983}, however we will not need to consider this effect in the present work.}.  Additionally, it is a simple matter to show that the moduli generated by the broken coordinate rotational symmetry $SO(3)_{\rm ext}$ are equivalent to the $(x,y)$ translations due to the so called inverse Higgs mechanism \cite{Ivanov:1975, Clark:2003, Low:2001bw, Nitta:2013mj}.  Figure 3 illustrates this equivalence.  Thus the four moduli generated from coordinate rotations and translations reduce to only one gapless mode.  We will extend this analysis to the quantization of the non-Abelian modulus fields appearing from the broken $SO(3)_{S+L}$.

\begin{figure}[h!]
\centering
\includegraphics[width=0.7\linewidth]{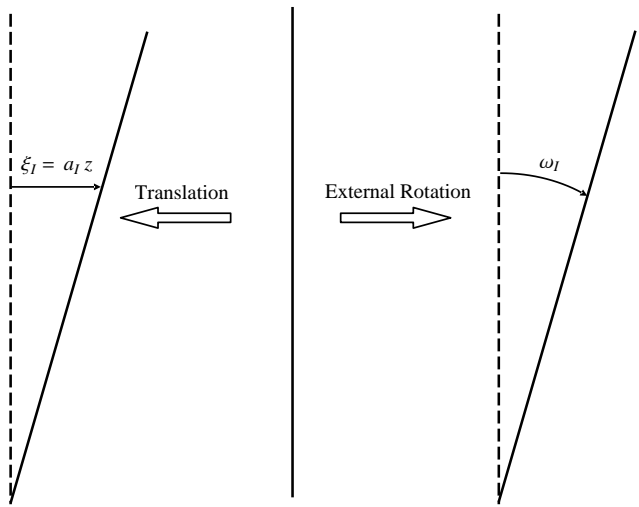}
\caption{The equivalence of infinitesimal external rotations with $z$-dependent translations is illustrated above.  Here the translation $\xi_I(z)$ is a linear function of $z$, and $\omega_I$ is the corresponding rotation angle. This effect is known as the inverse Higgs mechanism \cite{Ivanov:1975, Clark:2003, Low:2001bw, Nitta:2013mj}.}
\end{figure}

The organization of our presentation is as follows.  We will begin with a brief review of the Ginzburg-Landau description of superfluid $^3$He, and discuss the emergence of topological vortices in the bulk B phase.  The following section will present a classification of vortices, where we will determine the general form of the order parameter for an axially symmetric vortex.  In particular we will determine some specific requirements for the axially symmetric components to emerge from the trace-only solution.  In the following sections we will determine the static low energy theory describing the translational and non-Abelian modulus fields localized on the mass vortex with the modifications due to the symmetric components of the axially symmetric vortex.  Finally, we will characterize the precise number and type of quantized gapless modes appearing on the axial vortex by considering the Goldstone theorem in non-relativistic systems.  Although we will discuss the effects of quantization on the classically gapless modes, we will not discuss quantum corrections coming from loops in the effective potential.  For such an analysis we turn the reader to the discussion presented in \cite{Nitta:2013wca}.

\section{The Ginzburg-Landau description}
In this section we will review the Ginzburg-Landau description of condensed matter systems following the descriptions outlined in \cite{Volovik:2006a, Mineev:1998, Leggett:2006}.  The Ginzburg-Landau theory provides a macroscopic description of the order parameter of superfluid $^3$He near the critical temperature.  The theory can be derived from consideration of the microscopic BCS theory of the p-wave Cooper pairs of the $^3$He atoms resulting in a non-relativistic field theory of the order parameter.  As we discussed in the introduction the order parameter is a complex $3 \times 3$ matrix that transforms as a vector under both spin and orbital rotations \cite{Leggett:1975a,Thuneberg:1987a,Sauls:1981}
\begin{equation}
e_{\mu i} \rightarrow {\rm e}^{{\rm i} \psi}S_{\mu \nu}L_{i j}e_{\nu j},
\end{equation}
where $S_{\mu \nu}$ and $L_{ij}$ are spin and orbital rotations  respectively.  The symmetry group describing these transformations is 
\begin{equation}
G = U(1)_P \times SO(3)_S \times SO(3)_L
\end{equation}
where $U(1)_P$ is the group of phase rotations ${\rm e}^{{\rm i} \psi}$, and $SO(3)_{S,L}$ are the groups of spin and orbital rotations.  The most general (time-dependent) Ginzburg-Landau free energy containing this symmetry is given by \cite{Thuneberg:1987a,Volovik:1986a,Mermin:1973, Buchholtz:1977}
\begin{eqnarray}
&F_{\rm GL}=F_{\rm time}+F_{\rm grad}+V, \nonumber \\
&F_{\rm time}={\rm i} e_{\mu i}\partial_t e^{\star}_{\mu i}, \nonumber \\
&F_{\rm grad}=\gamma_1 \partial_i e_{\mu j} \partial_i e^{\star}_{\mu j}+\gamma_2 \partial_i e_{\mu i} \partial_j e^{\star}_{\mu j}+\gamma_3 \partial_i e_{\mu j} \partial_j e^{\star}_{\mu i}, \nonumber \\
&V=-\alpha e_{\mu i} e^{\star}_{\mu i}+\beta_1 e^{\star}_{\mu i} e^{\star}_{\mu i} e_{\nu j} e_{\nu j}+\beta_2 e^{\star}_{\mu i} e_{\mu i} e^{\star}_{\nu j} e_{\nu j} +\beta_3 e^{\star}_{\mu i} e^{\star}_{\nu i} e_{\mu j} e_{\nu j} \nonumber \\
&+\beta_4 e^{\star}_{\mu i} e_{\nu i} e^{\star}_{\nu j} e_{\mu j} +\beta_5 e^{\star}_{\mu i} e_{\nu i} e_{\nu j} e^{\star}_{\mu j},
\label{GLFE}
\end{eqnarray}
where $\gamma_i$, $\alpha$, and $\beta_i$ are phenomenological parameters whose values can be determined at zero pressure from the BCS theory \cite{Sauls:1981}
\begin{eqnarray}
&\alpha = \frac{N(0)}{3}\left(1-\frac{T}{T_c}\right), \nonumber \\
&-2\beta_1 = \beta_2 = \beta_3 = \beta_4 =  -\beta_5 = \frac{7N(0)\zeta(3)}{120(\pi T)^2}, \nonumber \\
&\gamma_1 = \gamma_2 = \gamma_3 = 7\zeta(3)N(0)\frac{v_F^2}{240(\pi T)^2},
\end{eqnarray}
where $N(0) = m^{\star}k_F/2\pi^2\hbar^2$.
Strong coupling corrections for the specific case of superfluid $^3$He have been determined in \cite{Choi:2007a}.  However, we quickly point out that in the present paper we will adjust the constants at our will depending on the particular features we wish to illustrate.  The time-dependent part $F_{\rm time}$ of (\ref{GLFE}) is typically discussed for non-equilibrium dynamics where quasi-static approximations are not valid (see for example \cite{Kopnin:1992} and \cite{Tang:1995}).

In this work we will mostly consider the case where $\gamma_{2,3} \rightarrow 0$ where we may distinguish internal and external orbital rotations $SO(3)_L \rightarrow SO(3)_{L_{\rm int}} \times SO(3)_{L_{\rm ext}}$.  This can be theoretically achieved for ultra-cold fermion p-wave pairs, however this requirement does not hold for general systems.  Thus our results below will be somewhat illustrative only, and should not necessarily be considered for precise measurements.

Minimizing the free energy (\ref{GLFE}) results in several different vacua, two of which are achieved physically in superfluid $^3$He.  These are the A-phase and the B-phase.  We will focus our attention on the B-phase vacuum, which is given by the retention of a spin-orbit locked $SO(3)$ symmetry
\begin{equation}
G \rightarrow H_B = SO(3)_{S+L}.
\end{equation}
The order parameter in the B-phase is given by
\begin{equation}
(e_B)_{\mu i}={\rm e}^{{\rm i} \psi}\Delta (R_0)_{\mu i}, \; \Delta = \frac{\alpha}{6\beta_{12}+2\beta_{345}},
\label{OPB}
\end{equation}
where $(R_0)_{\mu i}$ is a generic rotation matrix and the gap parameter $\Delta$ is determined by minimization of (\ref{GLFE}).  Here and throughout this analysis we will make use of the shorthand notation 
\begin{eqnarray}
&\gamma_{abc...} = \gamma_a + \gamma_b +\gamma_c + ..., \nonumber \\
&\beta_{abc...} = \beta_a + \beta_b + \beta_c + ...
\end{eqnarray}
We can see that the order parameter (\ref{OPB}) is invariant under simultaneous orbital $L_{ij}$ and spin $S_{\mu \nu}$ rotations satisfying $S = R_0 L R_0^T$.

In the B-phase the free energy (\ref{GLFE}) has a ground state degeneracy
\begin{equation}
G/H_B = U(1)_P \times SO(3)_{S-L}.
\label{DegeneracySpace}
\end{equation}
We may thus select the state with $(R_0)_{\mu i} = \delta_{\mu i}$ as our ground state.  Considering the first fundamental group of the degeneracy space (\ref{DegeneracySpace})
\begin{equation}
\pi_1(G/H_B) = \pi_1(U(1))+\pi_1(SO(3)) = \mathbb{Z} + \mathbb{Z}_2
\end{equation}
we see that the B-phase admits topologically stable mass vortices with integer topological charge $n \in \mathbb{Z}$.  Additionally, spin vortices with $\mathbb{Z}_2$ winding $\nu \in (0,1)$ also appear in the vacuum.  We will only consider the mass vortices with windings $n = \pm 1$ and $\nu = 0$ \cite{Mineev:1998}.

The precise form of the single vortex $(n=1)$ order parameter must be determined by minimization of the free energy (\ref{GLFE}).  For this purpose we will consider solutions of the form
\begin{equation}
e_{\mu i} = {\rm e}^{\rm i \phi}f(\vec{x}_\perp)\delta_{\mu i} + \varepsilon_{\mu i k} \chi_k(\vec{x}_\perp)+S_{\mu i}(\vec{x}_\perp), \; \Tr(S)=0,
\end{equation}
where $f$, $\chi_k$, and $S_{\mu i}$ are functions to be determined by minimization.  Here $\phi$ refers to the polar angle about the vortex axis.  To satisfy the asymptotic B phase condition we must require
\begin{equation}
f \rightarrow \Delta, \mbox{ as } |\vec{x}_\perp| \equiv r \rightarrow \infty.
\end{equation}
Additionally, $f$ must vanish as $r \rightarrow 0$ to satisfy the winding condition at the origin.  In addition, far from the vortex center $\vec{x}_\perp \rightarrow \infty$ both $\chi_i$ and $S_{\mu i}$ will be required to vanish to satisfy the B phase vacuum constraint.

\section{Symmetry classification of vortices and axially symmetric solutions}

\subsection{Symmerty structure of vortex solutions}
In this section we will discuss the topic of vortex classification by considering the symmetries broken by specific vortex solutions.  The results of this section will allow us to determine vortex solutions by reducing the number of components of the order parameter to those that satisfy invariance under unbroken symmetries.  Additionally, by simply considering the broken generators we will be able to determine the modulus fields emerging on the mass vortex, and draw general conclusions about their kinetic terms and interactions in a low energy effective theory.  The presentation of this section will follow closely the approaches given in \cite{Volovik:2006a,Salomaa:1985,Thuneberg:1987a}.

For the moment we will set $\gamma_{2,3} = 0$ and consider the complete symmetry group $\mathcal{G}$ of the free energy (\ref{GLFE}), which includes both the continuous symmetries from $G$ as well as the coordinate translational and the discrete symmetries.  In addition $\mathcal{G}$ includes the enhanced symmetry distinguishing internal and external orbital rotations $SO(3)_L \rightarrow SO(3)_{L_{\rm int}} \times SO(3)_{L_{\rm ext}}$.  We may split $\mathcal{G}$ into its continuous and discrete groups
\begin{equation}
\mathcal{G} = \mathcal{G}_{\rm cont} \times \mathcal{G}_{\rm dis},
\end{equation}
where
\begin{eqnarray}
&\mathcal{G}_{\rm cont} = U(1)_P \times SO(3)_S \times SO(3)_{L_{\rm int}} \times SO(3)_{ L_{\rm ext}} \times T \nonumber \\
&\mathcal{G}_{\rm dis} = \mathcal{T} \times P.
\end{eqnarray}
Here $T$ represents the translational symmetry, $SO(3)_{L_{\rm ext}}$ represents rotations of the coordinate.  The generator $L_{\rm int}$ will henceforth refer to the rotations of the orbital index of the order parameter.  The discrete symmetries $\mathcal{T}$ and $P$ represent time reversal and parity transformations.  The effect of these transformations on the order parameter $e_{\mu i}$ are given by the following
\begin{eqnarray}
&U_{\theta}={\rm e}^{{\rm i}\hat{I}\theta} \in U(1)_P, \mbox{ where } \;\hat{I}e_{\mu i} = e_{\mu i}, \; \hat{I}e_{\mu i}^{\star}=-e_{\mu i}^{\star}, \nonumber \\
&\hat{S}_{\beta}e_{\mu i}=-{\rm i}\varepsilon_{\beta \mu \nu}e_{\nu i}, \; {\rm e}^{{\rm i}\vec{\omega}\cdot \hat{S}} \in SO(3)_S \nonumber \\
&\hat{L}^{\rm int}_je_{\mu i} = -{\rm i}\varepsilon_{jik}e_{\mu k}, \; {\rm e}^{{\rm i}\vec{\omega} \cdot \hat{L}^{\rm int}} \in SO(3)_{L_{\rm int}}\nonumber \\
&\hat{L}^{\rm ext}_je_{\mu i} = -{\rm i}\varepsilon_{jlk}x_l\frac{\partial}{\partial x_k}e_{\mu i}, \; {\rm e}^{{\rm i}\vec{\omega} \cdot \hat{L}^{\rm ext}} \in SO(3)_{L_{\rm ext}}, \nonumber \\
&T_{\vec{\xi}_0} e_{\mu i}(\vec{x})=e_{\mu i}(\vec{x} - \vec{\xi}_0), \; T_{\vec{\xi}_0} \in T \nonumber \\
&\hat{\mathcal{T}}e_{\mu i} = e_{\mu i}^{\star}, \; \hat{\mathcal{T}} \in \mathcal{T}\nonumber \\
&\hat{P}e_{\mu i}(\vec{x})=-e_{\mu i}(-\vec{x}), \; \hat{P} \in P.
\label{Transformations}
\end{eqnarray}
In the first line of (\ref{Transformations}) $\theta$ represents the phase angle.  We will employ the shorthand notations
\begin{equation}
\hat{L} \equiv \hat{L}_{\rm int}+\hat{L}_{\rm ext}, \; \; \hat{J}_{\rm int} \equiv \hat{L}_{\rm int}+\hat{S}, \; \; \hat{J} \equiv \hat{L}_{\rm int}+\hat{L}_{\rm ext} + \hat{S}.
\end{equation}
In the B-phase vacuum state where we may select for our ground state
\begin{equation}
e_{\mu i} = {\rm e}^{{\rm i} \psi}\Delta \delta_{\mu i},
\label{TraceOnly}
\end{equation}
we can see that the preserved continuous symmetries from $\mathcal{G}$ are
\begin{equation}
\mathcal{H}_B = SO(3)_{J_{\rm int}} \times SO(3)_{L_{\rm ext}} \times T,
\end{equation}
as well as discrete symmetries given by the transformations
\begin{equation}
U_\psi \hat{\mathcal{T}} U_{-\psi}, \mbox{ and } \hat{P}U_\pi
\end{equation}

In the presence of a vortex in the B-phase the determination of the continuous symmetries preserved in the asymptotic limit is somewhat more subtle.  Far away from the vortex core the order parameter approaches the form
\begin{equation}
e_{\mu i} \rightarrow {\rm e}^{{\rm i} \phi} \Delta \delta_{\mu i}, \mbox{ as } r \rightarrow \infty.
\end{equation}
Clearly the asymptotic vacuum form retains the spin-orbit locking $SO(3)_{J_{\rm int}}$ symmetry as well as invariance under translations $T_{\vec{\xi}_0}$.  It is also not difficult to show that coordinate rotations about the $x$ and $y$ axis also leave the order parameter invariant in the asymptotic limit.  However, it is clear that the order parameter is not invariant under independent phase and external coordinate rotations about the $z$-axis.  Instead the order parameter is invariant under axial $U(1)_A$ transformations generated by a linear combination of the generators of phase and $z$-axis external coordinate rotations:
\begin{equation}
\hat{Q} = \hat{L}_z^{\rm ext}-\hat{I}, \; U_\delta = {\rm e}^{{\rm i}\delta\hat{Q}} \in U(1)_A.
\end{equation}
A specific example of an axial transformation generated by $\hat{Q}$ is shown in Figure 4.
\begin{figure}[h!]
\centering
\includegraphics[width=0.8\linewidth]{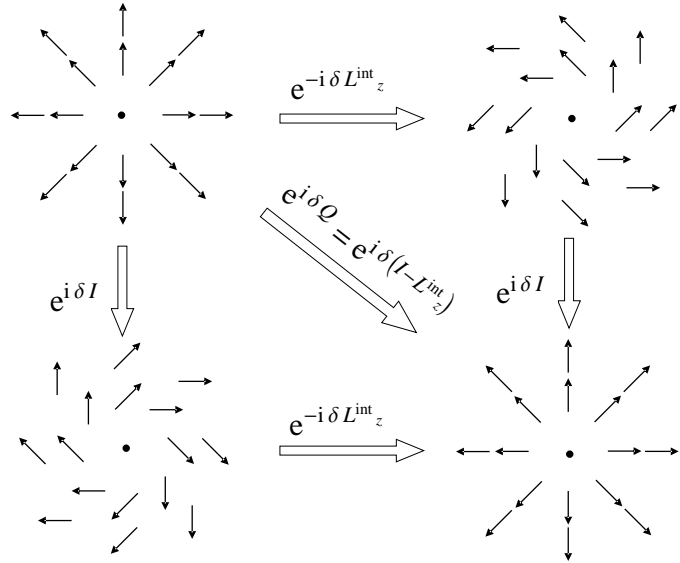}
\caption{The phase vectors of an axial $U(1)_A$ symmetric vortex solution are mapped onto the perpendicular plane to the vortex axis.  The two transformations generated by $\hat{L}_z^{\rm int}$ and $\hat{I}$ are shown in the upper right and lower left corner respectively.  Performing these transformations in succession by equal and opposite angle $\delta$ are known as axial $U(1)_A$ transformations generated by $\hat{Q}$ for which axially symmetric solutions are invariant as illustrated in the lower right corner.}
\end{figure}
The linear combinations of $\hat{Q}$ with $\hat{L}_x^{\rm ext}$ and $\hat{L}_y^{\rm ext}$ generate a coordinate-phase locked (axial) symmetry $SO(3)_{A+L_\perp^{\rm ext}}$, which contains $U(1)_A$ as a subgroup.
Thus summarizing the continuous symmetries of the B-phase with an $n=1$ vortex in the asymptotic limit
\begin{equation}
(\mathcal{H}_B)_{n=1} = SO(3)_{J_{\rm int}} \times SO(3)_{A+L_\perp^{\rm ext}} \times T
\label{CAsymVort}
\end{equation}
Additionally, in the asymptotic limit, the order parameter is invariant under the discrete transformations
\begin{equation}
\hat{P}_1 \equiv \hat{P}{\rm e}^{{\rm i} \pi \hat{I}}, \; \hat{P}_3 \equiv \hat{\mathcal{T}}{\rm e}^{{\rm i} \pi \hat{J}_x}.
\label{DAsymVort}
\end{equation}
Note that $\hat{J}_x$ generates rotations of both the internal and external degrees of freedom.

The degeneracy space associated with a particular mass vortex solutions is characterized by the possible symmetry breaking pattern of the continuous group (\ref{CAsymVort}).  Clearly the existence of a vortex core breaks translations in the $x$ and $y$ directions as well as rotations about any axis in the $xy$ plane.  Additionally, vortex solutions are invariant under translations in the $z$ direction.  Vortex solutions are thus classified according to their transformation properties under the group
\begin{equation}
U(1)_A \times SO(3)_{J_{\rm int}} \times (\mathbb{Z}_2 \times \mathbb{Z}_2)_{P_1 \times P_3}
\label{ClassGroup}
\end{equation}
whose continuous group elements are generated by $\hat{Q}$, $\hat{J}_{\rm int}$, with discrete group elements given in (\ref{DAsymVort}).

\subsection{Vortex solutions in the spherical tensor basis}
In searching for vortex solutions that minimize the free energy, we may consider forms of the order parameter that break some or all of the symmetries in (\ref{ClassGroup}).  In this light it is most convenient to expand the vortex solution in terms of eigenfunctions of $\hat{Q}$, $\hat{S}_z$, and $\hat{L}^{\rm int}_z$.  
\begin{eqnarray}
&e_{\mu i} = \sum_{\rho, \nu = \pm 1,0}\sum_nC_{\rho\nu, n}(r)\lambda_\mu^\rho \lambda_i^\nu {\rm e}^{{\rm i} n \phi}, \nonumber \\
&\hat{S}_z \lambda_\mu^\rho = \rho \lambda_\mu^\rho, \nonumber \\
&\hat{L}_z^{\rm int} \lambda_i^\nu = \nu \lambda_i^\nu,
\end{eqnarray}
where the $\hat{J}^{\rm int}_z$ eigenfunctions written in the cartesian $(x,y,z)$ basis as 
\begin{equation}
\lambda_{\alpha}^\pm = 1/\sqrt{2}(\hat{x}_\alpha \pm i \hat{y}_\alpha), \; \lambda_\alpha^0 = \hat{z}_\alpha.
\end{equation}

The most trivial example of a solution is a vortex that is axially $U(1)_A$ symmetric and invariant under $SO(3)_{J_{\rm int}}$
\begin{equation}
\hat{Q}e_{\mu i}  = \hat{J}_{\rm int}e_{\mu i} = 0.
\end{equation}
In this case the solution is constrained to $n = 1$, as well as $\rho + \nu = 0$, with $C_{+-} = C_{00} = C_{-+}  = f(r)$, and all other $C_{\rho \nu}$ vanish.  Additionally, the invariance under the $(\mathbb{Z}_2 \times \mathbb{Z}_2)_{P_1 \times P_3}$ requires $f(r)$ to be real.  This would represent the trace-only vortex solution 
\begin{equation}
e_{\mu i}^{t}(x,y) = {\rm e}^{{\rm i}\phi} f(r) \delta_{\mu i}.
\end{equation}

A more interesting ansatz occurs when we consider solutions that are invariant under a locked $U(1)_{A+J^{\rm int}_z}$ symmetry, which is a subgroup of the classification group (\ref{ClassGroup}) and is generated by
\begin{equation}
\hat{Q}'e_{\mu i} = 0, \mbox{ where } \hat{Q}' \equiv \hat{Q}+\hat{J}^{\rm int}_z= \hat{J}-\hat{I}.
\end{equation}
This results in the constraint $n+\rho + \nu = 1$, and thus we may express the amplitudes $C_{\rho\nu, n}$ as 
\begin{equation}
\sum_n C_{\rho\nu,n}=
\left(\begin{array}{ccc} 
C_{++}{\rm e}^{-{\rm i}\phi} & C_{+0} & C_{+-}{\rm e}^{{\rm i} \phi}\\
C_{0+} & C_{00}{\rm e}^{{\rm i} \phi} & C_{0-}{\rm e}^{2{\rm i}\phi} \\
C_{-+}{\rm e}^{{\rm i} \phi} & C_{-0}{\rm e}^{2{\rm i}\phi} & C_{--}{\rm e}^{3{\rm i}\phi} \\
\end{array}\right)
\label{Axisymmetric}
\end{equation}
The $U(1)_{A+J_z}$ vortices are further categorized by their transformations under the $(\mathbb{Z}_2 \times \mathbb{Z}_2)_{P_1 \times P_3}$.  This results in five subclasses of axisymmetric vortices.  We will consider the subclass given by
\begin{equation}
\hat{P}_1\hat{P}_3 e_{\mu i} =e_{\mu i}, 
\end{equation} 
which reduces to the constraint that the $C_{\rho\nu}$ amplitudes are real.

To identify the trace, symmetric, and anti-symmetric components of $e_{\mu i}$ for the axially symmetric solutions we rearrange the solution in terms of the irreducible multiplets in the $(J,J_z)$ basis.  Following this procedure may write the antisymmetric and symmetric-traceless parts of the order parameter $e_{\mu i}$ as follows
\begin{equation}
e_{\mu i} = {\rm e}^{{\rm i} \phi}f(r)\delta_{\mu i} + \varepsilon_{\mu i j}\chi_j(r,\phi)+S_{\mu i}(r,\phi),
\end{equation}
where we write the $\chi_i$ and $S_{\mu i}$ tensors in terms of the spherical tensor components $\chi_{m_J}$ and $s_{m_J}$,
\begin{equation}
\vec{\chi}(r,\phi) = {\rm i} \left( \chi_1\vec{\lambda}^+ +\chi_0 {\rm e}^{{\rm i}\phi}\vec{\lambda}^0+\chi_{-1}{\rm e}^{2{\rm}i\phi}\vec{\lambda}^-\right),
\end{equation}
and
\begin{eqnarray}
S_{\mu i} = &s_{2}{\rm e}^{-{\rm i}\phi}\lambda_{\left\{ \mu \vphantom{i} \right.}^{+\vphantom{+}} \lambda_{\left. \vphantom{\mu} i \right\}}^{+\vphantom{+}}+ s_{1}\lambda_{\left\{ \mu \vphantom{i} \right.}^{+\vphantom{0}} \lambda_{\left. \vphantom{\mu} i \right\}}^{0\vphantom{+}}\nonumber \\ 
&+s_{0}{\rm e}^{{\rm i}\phi}\left(\lambda_{\left\{ \mu \vphantom{i} \right.}^{+\vphantom{-}} \lambda_{\left. \vphantom{\mu} i \right\}}^{-\vphantom{+}}-2\lambda_{\mu \vphantom{i}}^{0\vphantom{0}} \lambda_{\vphantom{\mu} i}^{0\vphantom{0}} \right)  \nonumber \\
&+ s_{-1}{\rm e}^{2{\rm i}\phi}\lambda_{\left\{ \mu \vphantom{i} \right.}^{-\vphantom{0}} \lambda_{\left. \vphantom{\mu} i \right\}}^{0\vphantom{-}}+ s_{-2}{\rm e}^{3{\rm i}\phi}\lambda_{\left\{ \mu \vphantom{i} \right.}^{-\vphantom{-}} \lambda_{\left. \vphantom{\mu} i \right\}}^{-\vphantom{-}}.
\end{eqnarray}

The following relations between the $(J,J_z)$ and $(L^{\rm int}_z, S_z)$ bases are useful and easy to derive from the Clebsch-Gordon coefficients:
\begin{eqnarray}
& f = \frac{1}{3}(C_{+-}+C_{-+}+C_{00}), \nonumber \\
& \chi_{\pm 1} = \frac{1}{2\sqrt{2}}(C_{\pm 0}-C_{0 \pm}), \;  \chi_0 = \frac{1}{2}(C_{-+}-C_{+-}), \nonumber \\
& s_{\pm 2} =\frac{C_{\pm \pm}}{2}, \; s_{\pm 1} = \frac{1}{2\sqrt{2}}(C_{\pm 0}+C_{0 \pm}), \; s_0 = \frac{1}{6}(C_{+-}+C_{-+}-2C_{00}).
\end{eqnarray}

In the context of superfluid $^3$He-B the solutions (\ref{Axisymmetric}) are known as the $v$-vortices \cite{Salomaa:1985}, and are characterized by a core with mixed A-phase $(C_{+0})$ and ferromagnetic $\beta$-phase $(C_{0+})$ components.  In the following section we will argue that these components will typically arise spontaneously from the trace only solution (\ref{TraceOnly}) and thus the vortex will contain symmetric and anti-symmetric components resulting in additional gapless excitations of the vortex core.

\section{Emergence of off-diagonal components in the vortex core}

In a previous work we considered the conditions under which a single real antisymmetric component $\varepsilon_{\mu i k}\chi_k(r)$ of the vortex order parameter would spontaneously develop in the core from the initial trace-only solution (\ref{TraceOnly}).  To accomplish that task we employed the methods of \cite{Witten:1984eb} considered for superconducting strings.  Initially, the $\chi_k$ field is set to zero and the free energy (\ref{GLFE}) is minimized numerically by $f(r)$ (see Figure 5). 

\begin{figure}[h!]
\centering
\includegraphics{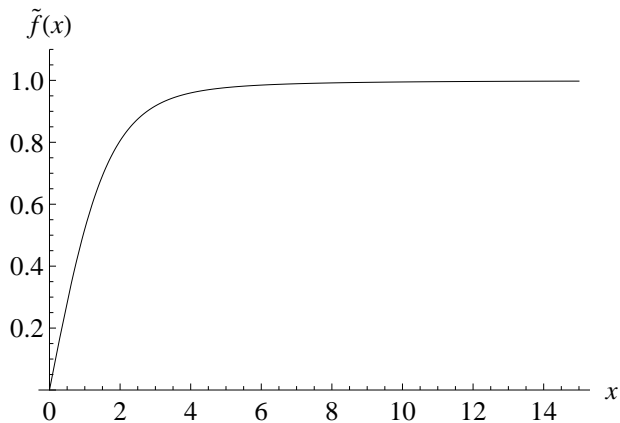}
\caption{The numerical solution $\tilde{f}(x) = f(x)/\Delta$ is plotted.  Here we have defined $x \equiv r \sqrt{\alpha/\gamma_1}$.  For $x \ll 1$ the solution follows the form $\tilde{f}(x) \sim 0.583x$.  In the opposite limit $x \gg 1$ the function $\tilde{f}(x) \rightarrow 1-1/2x^2+\mathcal{O}(x^{-4})$.}
\end{figure}

At this point a small $\chi_k(r)$ field is considered in the free energy (\ref{GLFE}) such that the quartic terms of $\chi_k$ may be neglected.  Since we consider $\gamma_{2,3} = 0$ we are free to set 
\begin{equation}
\vec{\chi}(r) \equiv (0,0,\chi(r)).
\end{equation}
Thus the $\chi(r)$ dependent part of the free energy is
\begin{equation}
F_{\chi} =\int r dr\left\{ i\chi\frac{\partial \chi}{\partial t}+\chi L_2 \chi +\mathcal{O}(\chi^4) \right\} ,
\label{Fchi}
\end{equation}
where $L_2$ is given given by:
\begin{eqnarray}
&L_2 = -2\gamma_1\frac{1}{r} \frac{\partial}{\partial r} \left(r\frac{\partial}{\partial r}\right)+V(r), \nonumber \\
&V(r) = 4(3\beta_2 + 2\beta_4)f^2(r) -2\alpha.
\label{Schrodinger}
\end{eqnarray}
Thus $L_2$ acts as a Schrodinger operator whose eigenvalues and eigenvectors are given by solving
\begin{equation}
L_2\chi_n = \omega_n \chi_n, \mbox{ where } \chi(r) = \sum_n a_n\chi_n(r).
\end{equation}
It can be shown that a negative eigenvalue $\omega_0$ will exist under the condition
\begin{equation}
\frac{1}{2} \le \frac{3\beta_2+2\beta_4}{6\beta_{12}+2\beta_{345}} \lesssim 0.76.
\label{Condition}
\end{equation}
If this condition is satisfied a non-zero $\chi(r)$ field will be energetically favorable compared to the trace-only solution (\ref{TraceOnly}).

We may take this analysis one step further by considering a more general ansatz for $\chi_k$, which satisfies the $U(1)_{A+J^{\rm int}_z}$ axial symmetry condition:
\begin{equation}
\vec{\chi}_A(r,\phi)={\rm i} \left(\chi_{1}(r)\vec{\lambda}^+ +\chi_{-1}(r){\rm e}^{2{\rm i} \phi}\vec{\lambda}^-\right).
\label{ComplexAnsatz}
\end{equation}
The stability is determined by considering the equations of motion linearized in $\chi_{\pm 1}$
\begin{eqnarray}
\fl &\frac{\gamma_1}{r} \frac{\partial}{\partial r}\left(r\frac{\partial \chi_1}{\partial r}\right)=((6\beta_2 +4\beta_4)f(r)^2 -\alpha)\chi_1+(6\beta_1-2\beta_4 +2\beta_{35})f(r)^2\chi_{-1}, \nonumber \\
\fl &\frac{\gamma_1}{r} \frac{\partial}{\partial r}\left(r\frac{\partial \chi_{-1}}{\partial r}\right)= \frac{4\chi_{-1}}{r^2}+((6\beta_2 +4\beta_4)f(r)^2 -\alpha)\chi_{-1}+(6\beta_1-2\beta_4 +2\beta_{35})f(r)^2\chi_1
\label{CoupledEquations}
\end{eqnarray}
Considering the case far from the vortex core a non-trivial solution to the coupled equations (\ref{CoupledEquations}) exists for any $\beta_i$ with the asymptotic condition
\begin{equation}
\chi_1(r) \simeq \chi_{-1}(r) \sim \frac{1}{r} + \mathcal{O}(r^{-3})
\end{equation}
Near the vortex core it is clear that $\chi_{-1}(r \rightarrow 0) \rightarrow 0$ due to the winding.  However, the field $\chi_1$ is not constrained at the origin by any winding and thus may develop a non-trivial value in the core.  This value must be determined by solving the full equations of motion.  Numerical solutions for $\chi_{\pm 1}(r)$ are shown in Figure 6 for a typical set of $\beta_i$ values.

Having demonstrated the spontaneous emergence of antisymmetric components $\chi_1$ and $\chi_{-1}$, we continue the analysis by considering symmetric-traceless components as perturbations on the antisymmetric solution.  For this we consider the symmetric ansatz
\begin{equation}
S_{\mu i}(r, \phi) = s_{1}\lambda_{\left\{ \mu \vphantom{i} \right.}^{+\vphantom{0}} \lambda_{\left. \vphantom{\mu} i \right\}}^{0\vphantom{+}} + s_{-1}{\rm e}^{2{\rm i}\phi}\lambda_{\left\{ \mu \vphantom{i} \right.}^{-\vphantom{0}} \lambda_{\left. \vphantom{\mu} i \right\}}^{0\vphantom{-}}.
\label{Axisymmetric}
\end{equation}
Considering the equations of motion linearized in $s_0$ and $s_2$ we arrive at
\begin{eqnarray}
&\frac{\gamma_1}{r} \frac{\partial}{\partial r}\left(r\frac{\partial s_1}{\partial r}\right)=2(\beta_5-\beta_3)\chi_1^3 +(\mbox{terms proportional to } s_{\pm 1}), \nonumber \\
&\frac{\gamma_1}{r} \frac{\partial}{\partial r}\left(r\frac{\partial s_{-1}}{\partial r}\right)=2(\beta_5-\beta_3)\chi_{-1}^3 +(\mbox{terms proportional to } s_{\pm 1}),
\label{CoupledSymmEquations}
\end{eqnarray}
where for our purposes it will only be necessary to consider the first terms on the righthand side of (\ref{CoupledSymmEquations}).  It is readily apparent that non-zero $\chi_{\pm 1}(r)$ fields act as sources for the symmetric-traceless fields $s_{\pm 1}(r)$.  Numerical solutions for $s_{\pm 1}(r)$ are shown in Figure 7.  Thus all four off diagonal fields $\chi_{\pm 1}$ and $s_{\pm 1}$ spontaneously develop from the trace-only solution (\ref{TraceOnly}).  The remaining four elements of the axial solution (numerically plotted in Figure 8) also arise spontaneously from (\ref{TraceOnly}) due to their couplings with the symmetric $s_{\pm 1}$ and antisymmetric $\chi_{\pm 1}$ solutions.  However, these are typically small and will have little impact on the low energy effective theory we discuss in the following sections.

\begin{figure}[h!]
\centering
\includegraphics{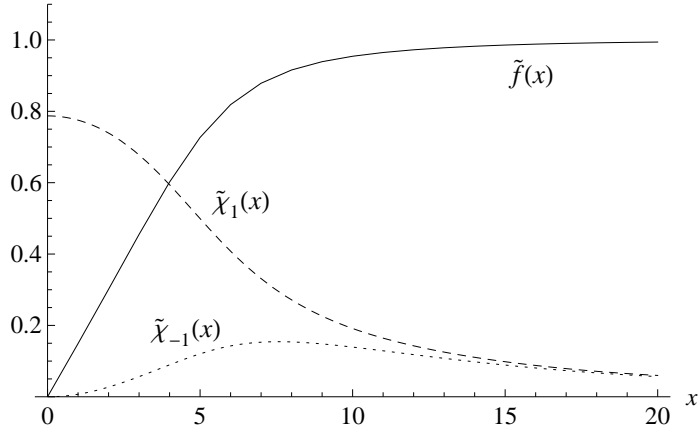}
\caption{The numerical solutions for $\tilde{f}(x) = f(x)/\Delta$, $\tilde{\chi}_1(x) = \chi_1(x)/\Delta$, and $\tilde{\chi}_{-1}(x) = \chi_{-1}(x)/\Delta$ are plotted, where again $x \equiv r \sqrt{\alpha/\gamma_1}$.  As $x \rightarrow \infty$ we have $\tilde{\chi}_1(x) \simeq \tilde{\chi}_{-1}(x) \rightarrow c/x$, where $c$ is a constant that must be determined by solving completely the vortex solution.}
\end{figure}

\begin{figure}[h!]
\centering
\includegraphics{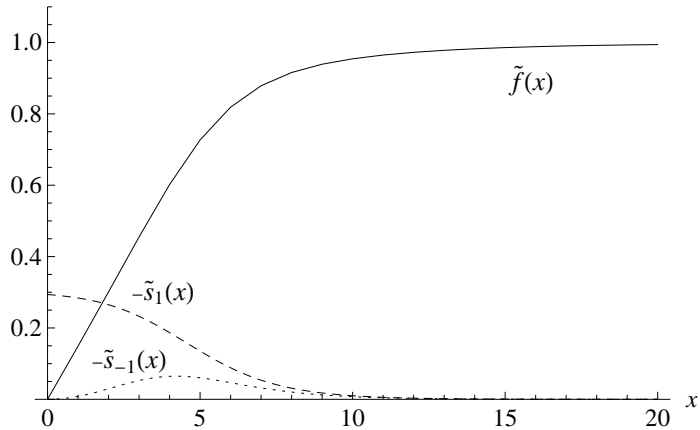}
\caption{The numerical solutions for $\tilde{f}(x) = f(x)/\Delta$, $\tilde{s}_1(x) = s_1(x)/\Delta$, and $\tilde{s}_{-1}(x) = s_{-1}(x)/\sqrt{2}\Delta$ are plotted, where again $x \equiv r \sqrt{\alpha/\gamma_1}$.}
\end{figure}

\begin{figure}[h!]
\centering
\includegraphics{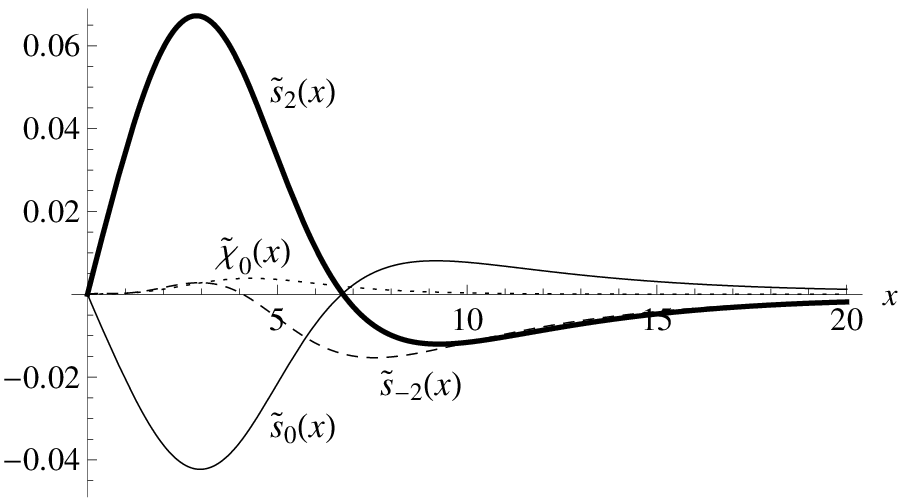}
\caption{The numerical solutions for $\tilde{s}_{\pm 2}(x) = s_{\pm 2}(x)/\Delta$, $\tilde{s}_0(x) = s_0(x)/\Delta$, and $\tilde{\chi}_0(x)=\chi_0(x)/\Delta$, where again $x \equiv r \sqrt{\alpha/\gamma_1}$.  These functions necessarily develop in response to the development of non-zero $\chi_{\pm 1}$ and $s_{\pm 1}$.  They are however typically small for most values of $\beta_i$.}
\end{figure}

\section{Broken symmetries and non-Abelian moduli localized on vortex}

Having established the spontaneous emergence of non-trace elements of the $n=1$ vortex order parameter we may proceed to address the effective field theory describing the gapless excitations of the vortex string.  At the classical level this can be accomplished by determining the broken symmetries of a particular vortex solution.  The number of broken symmetries determines the number of gapless moduli appearing in the low energy theory.  For relativistic systems this is precisely the number of quantized Goldstone modes appearing in the effective theory.  However, for non-relativistic systems, the Goldstone theorem is more subtle and care must be taken to determine how many gapless modes survive the quantization procedure.  In this section we will simply discuss the classical moduli determined from symmetry considerations and determine the classical effective theory.  In the following section we will consider quantization.

In section 4 we discussed that at large distances from the vortex core, the order parameter was required to be invariant under the continuous symmetry group (\ref{CAsymVort})
\begin{equation}
(\mathcal{H}_B)_{n=1} = SO(3)_{J_{\rm int}} \times SO(3)_{A+L_\perp^{\rm ext}} \times T
\end{equation}
The moduli generating gapless modes for a particular vortex solution are determined from the broken generators of the group (\ref{CAsymVort}).  In particular, all vortex lines break the translational symmetries in the $xy$ plane resulting in two Abelian moduli generating the Kelvin excitations.  Additionally, vortex solutions that break the axial $U(1)_A$ symmetry generate an additional Abelian mode.  We may also consider the coordinate rotations about the $x$ and $y$ axes, which would naively be expected to generate two additional moduli.  However, it is a simple matter to show that the coordinate rotations about the $x$ and $y$ directions are equivalent to a $z$-dependent translation in the $x-y$ plane.  Thus, there are a total of three potential Abelian moduli arising on the vortex string.  These are well studied in the context of vortices in superfluid $^4$He.  

We are however interested in the additional non-abelian modulus fields occurring from the breaking of the $SO(3)_{J_{\rm int}}$ symmetry in the vortex core.  There will be an additional two or three moduli appearing on the vortex depending on the symmetry breaking pattern of $SO(3)_{J_{\rm int}}$.  If the vortex solution retains a $U(1)$ symmetry from the breaking of $SO(3)_{J_{\rm int}}$ then there will appear two modes $\omega_x$ and $\omega_y$ living on the degeneracy space
\begin{equation}
SO(3)_{J_{\rm int}}/U(1)_{J^{\rm int}_z} \simeq S^2.
\end{equation}
An example of such a vortex solution is given by the case of an order parameter with a single real valued antisymmetric component $\vec{\chi}(r)=\chi(r)\hat{z}$ as discussed in at the beginning of the previous section.

If the vortex completely breaks $SO(3)_{J_{\rm int}}$, all three potential moduli $\omega_x$, $\omega_y$, and $\omega_z$ will appear in the effective theory living on the space
\begin{equation}
SO(3)_{J_{\rm int}}/1 \simeq S^3/Z_2.
\end{equation}
The axial vortices preserving the $U(1)_{A+J^{\rm int}_z}$ symmetry are characterized by a complete breaking of $SO(3)_{J_{\rm int}}$.  However, due to the preserved $U(1)_{A+J^{\rm int}_z}$ symmetry only two of the moduli from the broken $SO(3)_{J_{\rm int}}$ will be independent of the axial modulus $\delta$ from $U(1)_A$.  Table 1 summarizes the degeneracy space and associated modulus fields developing on particular vortex solutions.

\begin{table}
\caption{\label{label}A summary of degeneracy space and associated moduli for the vortex solutions when $\gamma_2,\gamma_3 =0$ considered in the previous section is shown.  The first column indicates the core type solution.  Columns 2-5 indicate the modulus fields emerging in the various solutions.  The degeneracy spaces in the sixth column are denoted with subscripts indicating the group associated with the degeneracy.  Additionally, we have defined $J_{\rm int} \equiv S+L_{\rm int}$.  The last column shows the total number of emerging moduli.}
\begin{indented}
\item[]\begin{tabular}{@{}lllllcc}
\br
Core Type & $\vec{\xi}_{\perp}$ & $\omega_{x,y}$ & $\omega_z$ & $\delta$ & Degeneracy Space  & Number of Moduli \\ 
\mr
$\vec{\chi}=0, \; f = f(r)$ & $\checkmark$ & $\times$ & $\times$ & $\times$ & $S^2_T$ & 2\\
$\vec{\chi}=0, \; f = f(x,y)$ & $\checkmark$ & $\times$ & $\times$ & $\checkmark$ & $S^1_A \times S^2_T$ & 3\\
$\vec{\chi}=\chi_z(r)\hat{z} \in \mathbb{R}^3$ & $\checkmark$  &  $\checkmark$  & $\times$ & $\checkmark$ & $S^2_{J_{\rm int}}\times S^1_A \times S^2_T$ & 5\\ 
Preserved $U(1)_{A+J_z}$ & $\checkmark$ & $\checkmark$ &  \multicolumn{2}{c}{$\omega_z \sim \delta^{\rm a}$} & $S^2_{J_{\rm int}+A} \times S^1_{J^{\rm int}_z+A_z} \times S_T^2$ & 5\\
Broken $U(1)_{A+J_z}$ & $\checkmark$ & $\checkmark$ & $\checkmark$ & $\checkmark$ & $(S^3/\mathbb{Z}^2)_{J_{\rm int}} \times S^1_A \times S^2_T$ & 6 \\
\br
\end{tabular}
\item[] $^{\rm a}$ We have noted the equivalence of $\delta$ and $\omega_z$ moduli for the A-phase core.
\end{indented}
\end{table}

For the case of axially asymmetric vortices the group $U(1)_A \times SO(3)_{J_{\rm int}}$ is completely broken, and thus all four moduli from the internal space will appear independently in the classical effective theory.  Physical examples include the double core vortex in superfluid $^3$He-B.  We will avoid further discussion of the axially asymmetric solutions of this form here.

Extending this analysis to the case of small but non-zero $\gamma_{2,3}$ it is readily seen from (\ref{GLFE}) that the associated gradient terms no longer have the separate $SO(3)_{L_{\rm int}}\times SO(3)_{L_{\rm ext}}$ from the internal and external rotations.  Instead the gradient terms preserve only the complete orbital rotations $SO(3)_{L^{\rm ext}+L^{\rm int}} = SO(3)_L$.  Thus due to the equivalence of coordinate rotations about the $x$ and $y$ axes to translations in the $xy$-plane we will find that only the translational moduli $\xi_{x,y}$ and the axial $U(1)_A$ modulus $\delta$ will be independent.  The moduli from the remaining generators will develop a mass gap proportional to $\gamma_{23}$.  In addition interactions between $\omega_{x,y}$, and $\xi_{x,y}$ will appear illustrating the equivalence of coordinate rotations and translations.

\section{Low energy effective field theory of gapless excitations}

Having established the existence of vortex solutions that spontaneously break the non-Abelian symmetry $SO(3)_{J_{\rm int}}$ in the vortex core we may write down the effective low energy theory of the gapless excitations arising from the broken symmetries.  In this section we will summarize the general procedure outlined in \cite{Nitta:2013mj} for determining this low energy theory by considering perturbations of the vortex line.  We will find that at least classically, the combinations of perturbations corresponding to the broken generators will contain no mass gap.  Upon quantization we will observe the number of gapless modes emerging from the classical modulus fields.  In the following section we will consider the specific solutions from the previous section and use the procedure discussed here to determine the quantized low energy theory of the gapless excitations.

We begin by considering fluctuations of the vortex line given by
\begin{equation}
e_{\mu i}(\vec{x}_{\perp}) = e^{\rm vort}_{\mu i}(\vec{x}_{\perp})+\delta e_{\mu i}(\vec{x}_\perp,z,t).
\label{Fluctuations}
\end{equation}
Inserting (\ref{Fluctuations}) into the free energy (\ref{GLFE}) and expanding to second order in $\delta e_{\mu i}$ we obtain upon integrating by parts in the spatial gradients
\begin{eqnarray}
&\delta^2 F_{\rm GL} = i \delta e_{\mu i} \partial_t \delta e_{\mu i}^{\star}+\delta e_{\mu i} L_{ij,\mu\nu}\delta e_{\nu j}^{\star}, \nonumber \\
& L_{ij,\mu\nu} = -\gamma_1\delta_{ij}\delta_{\mu\nu}\vec{\partial}^2-\gamma_{23}\delta_{\mu\nu}\partial_i \partial_j+(\partial_{e_{\mu i}^{\star} }\partial_{e_{\nu j}}V),
\label{EFT}
\end{eqnarray}
where we have consolidated the spatial gradient and potential terms into $L_{ij, \mu \nu}$.  At this point we consider an adiabatic mode expansion for $\delta e_{\mu i}(\vec{x}_\perp,z,t)$
\begin{equation}
\delta e_{\mu i}(\vec{x}_\perp,z,t) = \sum_n c_n(t,z) e_{\mu i}^{(n)}(\vec{x}_\perp)
\label{Expansion}
\end{equation}
where the functions $e^{(n)}_{\mu i}(\vec{x}_\perp)$ are eigenfunctions of $L_{ij,\mu \nu}$
\begin{equation}
L_{ij,\mu \nu}(\vec{x}_\perp)e^{(n)}_{\nu j}(\vec{x}_\perp)=E^{(n)}e^{(n)}_{\mu i}(\vec{x}_\perp).
\end{equation}
In the low energy approximation we may restrict the expansion (\ref{Expansion}) to eigenfunctions $e^{(n)}_{\mu i}(\vec{x}_\perp)$ for which $E^{(n)}$ is small compared to the free energy density of the unperturbed vortex solution.

For the current problem we are interested specifically in the zero-modes for which $E^{(n)} = 0$.  These modes are generated by the non-trivial symmetry transformations of $e^{\rm vort}_{\mu i}(\vec{x}_\perp)$ that leave free energy invariant.  For each broken generator we parameterize the associated transformation of the vortex solution by a modulus $m^a$, and consider the family of equivalent vortex solutions $e^{\rm vort}_{\mu i}(\vec{x}_\perp, m^a)$ and define the ground state solution as $e^{\rm vort}_{\mu i}(\vec{x}_\perp, 0) \equiv e^{\rm vort}_{\mu i}(\vec{x}_\perp)$.  The gapless fluctuations are thus generated by the modulus fields $m^a$ varying in space and time.  Specifically for the adiabatic approximation $m^a = m^a(z,t)$ and the gapless fluctuations are given by
\begin{equation}
\delta e_{\mu i}(\vec{x}_\perp, z,t)_{E=0}=\sum_a m^a(z,t) \frac{\partial}{\partial m^a}e^{\rm vort}_{\mu i}(\vec{x}_\perp, m^b).
\label{GaplessFluctuations}
\end{equation}

Inserting (\ref{GaplessFluctuations}) into (\ref{EFT}) and integrating over $\vec{x}_\perp$ we arrive at a low energy effective field theory of gapless excitations localized on the mass vortex.  If the moduli we consider are strictly gapless we will find the free energy can be written as
\begin{eqnarray}
&F_{\rm eff}=iG^t_{ab}(m)m^a \partial_t m^b+G^z_{ab}(m)\partial_z m^a \partial_z m^b, \nonumber \\
&G^{t,z}_{ab}(m) = \int d^2 \vec{x}_{\perp}\frac{\partial e_{\mu i}^{\rm vort}(\vec{x}_{\perp},m)}{\partial m^a} \frac{\partial e_{\nu j}^{{\rm vort} \star}(\vec{x}_{\perp},m)}{\partial m^b} 
\label{GeneralLowE}
\end{eqnarray}
where $G_{ab}$ is a function of the modulus fields $m^a$.  We hasten to point out that the functions $G^{t,z}_{ab}(m)$ in (\ref{GeneralLowE}) are symbolic in the sense that their particular form depends on how the indices $i,j,\mu$, and $\nu$ are paired.

The second term of $F_{\rm eff}$ in (\ref{GeneralLowE}) describes the energy expense from fluctuations of the moduli from the broken generators.  To study the classical theory this is all that is required.  The first term is the dynamical term, which must be considered to determine the number of independent gapless propagating excitations emerging from the modulus fields.  For Lorentz invariant systems the corresponding dynamical term (which would include two time derivatives) would show that the number of quantized gapless modes is precisely given by the number of broken generators.  However, for a non-relativistic system, the Goldstone theorem is more subtle.  The simplest example of a non-relativistic system includes two modulus fields (such as the translational moduli) $m^1$ and $m^2$ where $G^t_{ab} \propto \varepsilon_{ab}$.  In this case the time dependent part of the free energy assumes the form
\begin{equation}
F_t = {\rm constant} \times  (m_1 \dot{m}_2 - m_2 \dot{m}_1).
\end{equation}
In this case the modulus fields are the conjugate momenta of each other and thus form a single Goldstone mode with quadratic dispersion, which we will refer to as a type A mode.  

On the other hand if $G^t_{ab} \propto \delta_{ab}$ the diagonal terms proportional to $m_a \dot{m}_a$ in the effective free energy are total derivatives at quadratic order and thus are non-dynamical.  In such cases where $F_t \propto m_a \dot{m}_a$  a more careful consideration of the interactions of $m_a$ with non-zero modes is required.  Typically this results in an effective theory of ``relativistic" Goldstone modes with linear dispersion.  The modes are known as Bogoliubov modes, which we will call type B.  Both type A and B Goldstone modes appear in the effective theories of the gapless excitations on mass vortices.

We hasten to point out that the Bogoliubov modes we discuss cannot always be considered as low energy excitations as their dispersion relations depend on the interaction with non-zero mode (gapped) fluctuations.  If the gap parameter associated with the non-zero mode is high, the propagation velocity of the Bogoliubov mode will be large, and the excitation will disappear from the spectrum.  In what follows below, we will assume that the non-zero modes have a small enough gap for the associated Bogoliubov mode to be considered in the low energy spectrum.

In the axially $U(1)_{A+J^{\rm int}_z}$ symmetric cases considered in the next section we will find that although the classical low energy theory predicts the existence of two separate translational moduli as well as three moduli from the internal symmetry breaking $SO(3)_{J_{\rm int}} \times U(1)_A \rightarrow U(1)_{A+J^{\rm int}_z}$, the quantized theory will show only one type A translational mode, one $U(1)$ Abelian type B mode, and one non-Abelian type A mode.  The reduction of two translational moduli to one quantized mode is of course well understood from the study of Kelvin excitations of vortices in superfluid $^4$He.  The $U(1)$ mode will follow from the breaking of the axial $U(1)_A$ symmetry.  The remaining two independent moduli $\omega_{x,y}$ will form a single quantized non-Abelian mode. For an axially asymmetric vortex completely breaking $SO(3)_{J_{\rm int}} \times U(1)_A$ an additional mode from $U(1)_{J_z^{\rm int}}$ would appear.

Before proceeding to a detailed discussion axially symmetric vortex excitations we wish to illustrate this method in detail by applying it to the case of a vortex with real vector $\vec{\chi} = \chi(r)\hat{z}$ and $S_{\mu i } \equiv 0$ considered in the previous section.  This ansatz is not axially $U(1)_{A+J^{\rm int}_z}$ symmetric, so we do not make any initial claims about the number and type of quantized modes in this case.  Applying general translations, rotations, and $U(1)_A$ transformations the moduli appear in
\begin{eqnarray}
&e_{\mu i}(\vec{x}_{\perp}) \rightarrow e_{\mu i}(\vec{x}_{\perp}-\vec{\xi}_{\perp}), \; \; \xi \in T \nonumber \\
&\chi^i(r) \rightarrow R_{ij}(\vec{\omega})\chi^j(r), \; \;R_{ij} \in SO(3)_{J_{\rm int}} \nonumber \\
&\chi^i(r) \rightarrow {\rm e}^{\rm i \delta}\chi^i(r), \; \;{\rm e}^{\rm i \delta} \in U(1)_A,
\label{ModuliTrans}
\end{eqnarray}
where $\vec{\xi}_{\perp}$, $\vec{\omega}$, and $\delta$ are functions of $z$ and $t$.  Additionally, it will be particularly convenient to consider the rotational moduli in the form
\begin{equation}
\chi^i = R_{ij}(\vec{\omega})\chi^j \equiv S^i(t,z)\chi(r), \; \; |S|^2  = 1,
\end{equation}
and consider the real modulus fields $\vec{S}(t,z)$ instead of $\vec{\omega}(t,z)$.  

Inserting this solution into (\ref{GLFE}) and integrating over $x$ and $y$ we arrive at the following low energy theory
\begin{eqnarray}
F_{\rm eff} = F_{\rm trans}+F_{O(3)}+F_{U(1)}, \nonumber \\
F_{\rm trans}=\frac{T_1}{2\gamma_1}\epsilon_{ab}\xi_a \partial_t \xi_b +\frac{T_2}{2}\partial_z  \vec{\xi}_{\perp}\cdot \partial_z \vec{\xi}_{\perp}, \nonumber \\
F_{U(1)} =\frac{1}{2g_1^2\gamma_1}\dot{\delta}^2+ \frac{1}{2g_2^2}\partial_z \delta \partial_z \delta, \nonumber \\
F_{O(3)} = \frac{1}{2g_2^2}\partial_z \vec{S} \cdot \partial_z \vec{S}, \; |S|^2 \equiv 1,
\label{EFTOriginal}
\end{eqnarray}
where $\epsilon_{ab}$ is the $2\times 2$ antisymmetric matrix.  The couplings $T_{1,2}$ and $g_{1,2}^2$ are determined from the integration over $\vec{x}_{\perp}$:
\begin{eqnarray}
\frac{T_{1,2}}{2} \sim \int d^2\vec{x}_{\perp} 3\gamma_1\frac{f^2}{r^2} \rightarrow \gamma_1\Delta^2 \log \left(\frac{\alpha R^2}{\gamma_1}\right), \\
\frac{1}{g_{1,2}^2} \sim  \int d^2\vec{x}_{\perp}\gamma_1\chi^2 \rightarrow  \frac{\gamma_1^2}{2\beta_{12}+\beta_{345}},
\label{Feff}
\end{eqnarray}
The low energy theory derived here shows the emergence of an $O(3)$ sigma model very similar in form to the low energy theory of gapless excitations of ANO strings in Yang-Mills theories.  For comparison see \cite{Monin:2013kza}.  Classically the low energy theory contains five gapless moduli from $\xi_{x,y}$, $S_{1,2}$, and $\delta$.

The $U(1)_A$ modulus $\delta$ appears in (\ref{EFTOriginal}) with a time derivative term at quadratic order.  This term can be derived by considering the fluctuations of the magnitude of the $\chi$ field in the vortex core.  Considering these fluctuations $\chi(x,y) \rightarrow \chi(x,y) + h(x,y,z,t)$ to quadratic order in $h$ in the free energy we arrive at a term linear in $h$ of the form
\begin{equation}
F \supset 2\chi(r)h\dot{\delta}.
\end{equation}
Integrating out the massive field $h$ and performing the integrations over the perpendicular coordinates $(x,y)$ we arrive at the quadratic time derivative term for $\delta$ in (\ref{EFTOriginal}).  We note that the $F_{O(3)}$ part of (\ref{EFTOriginal}) contains no time derivative terms since these terms only appear as total derivatives (to quadratic order in $\vec{S}$) and thus do not contribute to the low energy effective theory.  Naively it might be expected that the coupling of $\vec{S}$ to non-zero modes would produce quadratic time derivative terms of the form $\dot{S}_i \dot{S}_j$ in the effective free energy similar to the case of $\delta$.  However, a careful analysis shows that this is not the case.  Instead when we consider fluctuations of the form $\chi_i \rightarrow \chi(r) S_i(z,t) + h_i(z,t)$ the fluctuating non-zero modes $h_i$ coupling to $S_i$  appear in the effective theory as
\begin{equation}
F_{h_i} \supset 2{\rm i}\chi(r)\left(S_i \partial_t h_i  +h_i\partial_t S_i\right)+\mathcal{O}(h_i^2),
\end{equation}
which is a total derivative at linear order in $h_i$ and thus cannot contribute to the dynamical degrees of freedom.  This result follows from the requirement of time reversal symmetry.  If we relax the constraint that $\vec{\chi}$ be a real vector then we may consider complex fluctuations $h_i \in \mathbb{C}^3$ in which case the time derivative terms of $S_i$ will no longer be total derivatives, and two propagating non-Abelian modes may appear as Bogoliubov modes in form
\begin{equation}
F_{\dot{S}_i} = \frac{1}{2g_S^2\gamma_1}\partial_t\vec{S}\cdot \partial_t\vec{S}, \; \mbox{ with } \frac{1}{g_S^2} \sim \frac{\gamma_1^2}{2\beta_2-2\beta_1+\beta_4}.
\label{NonAbelianEFTModes}
\end{equation}

Summarizing the analysis the quantized effective theory (\ref{EFTOriginal}) describes the propagation of a translational Kelvin mode with quadratic dispersion and an axial $U(1)_A$ mode $\delta$ with linear dispersion along with potentially two additional Non-Abelian Bogoliubov modes from (\ref{NonAbelianEFTModes}).  The Kelvin and $U(1)_A$ modes appear on vortex lines of superfluid $^4$He and are well studied in that system.  The Non-Abelian modes are of course exclusive to systems with tensorial order parameter.

At this point we switch on a small but non-zero $\gamma_{23}$.  We will assume that $\gamma_{23}$ is small enough that we may neglect the corrections to the vortex solutions of $f(r)$ and $\chi(r)$, as well as the constants $T_{1,2}$ and $g_{1,2}^2$.  Aside from these uninteresting numerical corrections (\ref{Feff}) remains of the same form, however there are additional terms representing the breaking of the $SO(3)_{J_{\rm int}} \times SO(3)_{\rm L_{\rm ext}}$ symmetries.  They appear as follows
\begin{equation}
F_{\rm eff} \rightarrow F_{\rm eff}-\Delta F_{\rm eff},
\end{equation}
where $\Delta F_{\rm eff}$ represents symmetry breaking terms, which in this case are given by
\begin{eqnarray}
\Delta F_{\rm eff} = M^2(\vec{S}_{\perp}-S^3\partial_z \vec{\xi}_{\perp})^2 +\frac{\varepsilon}{2g^2}\left\{(S^3 \partial_z \delta)^2+(\partial_z S^3)^2\right\}.
\label{ModFeff}
\end{eqnarray}
Here $\varepsilon \sim \gamma_{23}/\gamma_1$, and $M$ represents a mass gap parameter given by
\begin{equation}
M^2 \sim \int d^2\vec{x}_{\perp}\gamma_{23} (\partial_{\perp} \chi)^2 \rightarrow \frac{\gamma_{23}\alpha}{2(2\beta_{12}+\beta_{345})}.
\end{equation}
In this form $M$ is known as the ``twisted mass" \cite{Alvarez:1983, Gates:1984}.  

For the case that $M^2 > 0$ the free energy will be minimized at $\vec{S}_\perp = 0$ and thus both moduli $S_{1,2}$ will no longer be gapless.  If $M^2 < 0$ $|\vec{S}_\perp| = 1$ and the effective theory will retain one non-Abelian gapless modulus.  In both cases the Kelvin and axial modes will remain gapless.

\section{Gapless modes of axially symmetric vortices}

At this point we may consider the low energy effective theory emerging on the $U(1)_{A+J^{\rm int}_z}$ axially symmetric vortex solutions, which present a richer, but more involved analysis.  As mentioned in the previous sections in the case that $\gamma_{23} = 0$ we expect to observe five modulus fields for the classical theory that will be reduced to one translational and one axial $U(1)_A$ mode, along with additional non-Abelian modes following the quantization.

We proceed by writing the axial vortex solution in the $(\hat{Q},\hat{S}_z,\hat{L}^{\rm int}_z)$ eigenbasis and performing translational, rotational, and $U(1)_A$ transformations
\begin{eqnarray}
\fl e_{\mu i} = \sum_{\rho,\nu}C_{\rho \nu}(r)\lambda_\mu^\rho& \lambda_i^\nu {\rm e}^{\rm i(1-\rho -\nu)\phi} \rightarrow \sum_{\rho,\nu}C_{\rho \nu}(r')\lambda_\mu^\rho(\vec{\omega}) \lambda_i^\nu(\vec{\omega}) {\rm e}^{\rm i(1-\rho -\nu)\phi'}{\rm e}^{\rm i(\rho +\nu)\delta}, \nonumber \\
&\lambda^{\rho}_a(\vec{\omega}) = R_{ab}(\vec{\omega})\lambda^\rho_b,
\label{ModDecomp}
\end{eqnarray}
where $(r',\phi')$ are the new coordinates generated by the translations $\vec{\xi}_\perp(z,t)$, and $R_{ab}(\vec{\omega})$ is a $SO(3)_{J_{\rm int}}$ rotation given as a function of the moduli $\vec{\omega}(z,t)$.  Henceforth, $\lambda^\rho_a$ will refer to the $\vec{\omega}$-dependent $\lambda^\rho_a(\vec{\omega})$.  Additionally, the $U(1)_A$ modulus $\delta$ is a function of $z$ and $t$.  Decomposing (\ref{ModDecomp}) into the trace, antisymmetric, and symmetric parts and converting to the $(J^{\rm int}, J^{\rm int}_z)$-basis we find
\begin{eqnarray}
e_{\mu i}(r,\phi,&z,t)  =  f{\rm e}^{{\rm i}\phi}\delta_{\mu i} \nonumber \\
&+{\rm i}\varepsilon_{\mu i k}\left(\chi_1\lambda_k^+{\rm e}^{{\rm i}\delta}+\chi_0{\rm e}^{{\rm i}\phi}\lambda_k^0+\chi_{-1} {\rm e}^{2{\rm i}\phi}\lambda_k^-{\rm e}^{-{\rm i}\delta} \right)\nonumber \\
& + s_{2}{\rm e}^{-{\rm i}\phi}\lambda_{\left\{ \mu \vphantom{i} \right.}^{+\vphantom{+}} \lambda_{\left. \vphantom{\mu} i \right\}}^{+\vphantom{+}}{\rm e}^{2{\rm i}\delta}+ s_{1}\lambda_{\left\{ \mu \vphantom{i} \right.}^{+\vphantom{0}} \lambda_{\left. \vphantom{\mu} i \right\}}^{0\vphantom{+}}{\rm e}^{{\rm i}\delta} \nonumber \\
&+ s_{0}{\rm e}^{{\rm i}\phi}\left(\lambda_{\left\{ \mu \vphantom{i} \right.}^{+\vphantom{-}} \lambda_{\left. \vphantom{\mu} i \right\}}^{-\vphantom{+}}-2\lambda_{\mu \vphantom{i}}^{0\vphantom{0}} \lambda_{\vphantom{\mu} i}^{0\vphantom{0}} \right) \nonumber \\
& + s_{-1}{\rm e}^{2{\rm i}\phi}\lambda_{\left\{ \mu \vphantom{i} \right.}^{-\vphantom{0}} \lambda_{\left. \vphantom{\mu} i \right\}}^{0\vphantom{-}}{\rm e}^{-{\rm i}\delta}+ s_{-2}{\rm e}^{3{\rm i}\phi}\lambda_{\left\{ \mu \vphantom{i} \right.}^{-\vphantom{-}} \lambda_{\left. \vphantom{\mu} i \right\}}^{-\vphantom{-}}{\rm e}^{-2{\rm i}\delta}.
\label{CartDecomp}
\end{eqnarray}
We will also neglect $s_{\pm 2,0}$ and $\chi_0$ in (\ref{CartDecomp}) as these components are typically much smaller than $f$, $\chi_{\pm 1}$ and $s_{\pm 1}$, and contribute only small corrections to the low energy effective theory.  Additionally, these contributions will not affect the number of quantized modes appearing in the theory.

Proceeding with these assumptions as well as the requirement that $\gamma_{23} = 0$ we may apply the methods of the previous section to arrive at the following low energy effective theory
\begin{eqnarray}
F_{\rm eff}=&\frac{T_1}{2}\varepsilon_{ab}\xi_a \partial_t\xi_b+\frac{T_2}{2}\partial_z \vec{\xi}\cdot \partial_z \vec{\xi} \nonumber \\
&+\frac{\rm i}{2g_1^2}\vec{\lambda}^+ \cdot \partial_t \vec{\lambda}^-+\frac{1}{2g_2^2}|\vec{\lambda}^- \cdot(\partial_t \vec{\lambda}^++i\vec{\lambda}^+\partial_t \delta)|^2  \nonumber \\
&+\frac{1}{2g_3^2}|\partial_z\vec{\lambda}^++i\vec{\lambda}^+\partial_z\delta|^2 - \frac{1}{2g_4^2}|\vec{\lambda}^- \cdot(\partial_z \vec{\lambda}^++i\vec{\lambda}^+\partial_z \delta)|^2,
\label{FeffAxial}
\end{eqnarray}
where again $\epsilon_{ab}$ is the $2 \times 2$ antisymmetric symbol and $T_{1,2}$ and $1/2g^2_{1,2,3,4}$ are given by
\begin{eqnarray}
\frac{T_{1,2}}{2} \sim \int d^2\vec{x}_{\perp} 3\gamma_1\frac{f^2}{r^2} \rightarrow \gamma_1\Delta^2 \log \left(\frac{\alpha R^2}{\gamma_1}\right), \\
\frac{1}{g_{1,3}^2} \sim  \int d^2\vec{x}_{\perp}\gamma_1(\chi_1^2+\chi_{-1}^2+s_1^2+s_{-1}^2) \rightarrow  \frac{\gamma_1^2}{\alpha}\Delta^2\log\left(\frac{\alpha R^2}{\gamma_1}\right), \\
\frac{1}{g_{2}^2} \sim  \int d^2\vec{x}_{\perp}\gamma_1(\chi_1^2-\chi_{-1}^2+s_1^2-s_{-1}^2) \rightarrow  \frac{\gamma_1^2}{4\beta_{2}+2\beta_{4}+\beta_{35}},\\
\frac{1}{g_{4}^2} \sim  \int d^2\vec{x}_{\perp}\gamma_1(s_1^2 +s_{-1}^2)\rightarrow  \frac{\gamma_1^2}{4\beta_{2}+2\beta_{4}+\beta_{35}}.
\label{ConstFeffAxial}
\end{eqnarray}
The interesting features to point out from (\ref{FeffAxial}) include the last two lines showing the $U(1)_{A+J_z}$ symmetry respected by the vortex solutions.  Thus as expected only five independent moduli from the broken translational and internal non-Abelian generators.  Additionally, the expected single Kelvin mode emerging from the dynamical term has appeared just as in the previous examples.  

The first term in the second line describing the time dependence of the non-Abelian modes is new.  In the limit of small oscillations we may consider rotations to the first order in $\vec{\omega}(z,t)$ and consider
\begin{equation}
\lambda_i^\rho(\vec{\omega})=R_{ij}(\vec{\omega})(\lambda_0)^\rho_j \simeq (\lambda_0)^\rho_i -\omega_j\epsilon_{ijk}(\lambda_0)_k^\rho,
\end{equation}
which gives the following result for the time dependent term
\begin{equation}
{\rm i}\vec{\lambda}^+ \cdot \partial \vec{\lambda}^- \simeq \omega_x \dot{\omega}_y-\omega_y \dot{\omega}_x.
\end{equation}
Thus we see that only one independent mode emerges from the modulus fields after quantization as the fields $\omega_{x,y}$ are the conjugate momenta of each other.  This mode is a type A Goldstone mode with quadratic dispersion relation.  Additionally the second term in the second line of (\ref{FeffAxial}) also includes a term quadratic in the time derivative of $\delta$ implying that the $U(1)_A$ Bogoliubov mode appears due to the coupling of $\delta$ with the massive oscillations of $\chi_{\pm 1}$ and $s_{\pm 1}$ as in the previous section.  After integrating these oscillations out we arrive at the corresponding term in (\ref{FeffAxial}).  It is clear that this term also preserves the $U(1)_{A+J^{\rm int}_z}$ symmetry.

For completeness we consider the additional gradient terms in the free energy (\ref{GLFE}) when $\gamma_{23} \neq 0$.  For the purposes of illustration we will omit the contributing terms from $s_{\pm 1}(r)$ as they are notationally complex but do not introduce any additional interesting effects.  As before we assume that the $\gamma_{23}$ corrections have negligible effect on the functions $f(r)$ and $\chi_{\pm 1}(r)$.  Thus we find the following additional terms to $F_{\rm eff} \rightarrow F_{\rm eff}-\Delta F_{\rm eff}$:
\begin{eqnarray}
\fl \Delta F_{\rm eff} \supset  M^2|\vec{\lambda}_{\perp}^+& -\lambda_3^+\partial_z \vec{\xi}_{\perp}|^2+\left\{M^2_{ab}\left(\lambda^+_a-\partial_z \xi_a \lambda^+_3\right)\left(\lambda^+_b-\partial_z \xi_b \lambda^+_3\right)e^{2i\delta} +{\rm h.c.} \right\} \nonumber \\
\fl &+\frac{\varepsilon}{2g^2}|\partial_z \lambda^+_3 + i \lambda^+_3 \partial_z \delta|^2,
\label{ModEffAxial}
\end{eqnarray}
where $\varepsilon = \gamma_{23}/\gamma_1$ and
\begin{equation}
M^2 \sim \int d^2\vec{x}_{\perp}\gamma_{23} (\partial_{\perp} \chi_{\pm 1})^2 \rightarrow \gamma_{23} \Delta^2\times C(\beta_i),
\end{equation}
where $C(\beta_i)$ is a constant determined by the $\beta_i$ parameters.  $M^2_{ab}$ represents a symmetric mass matrix illustrating the crossing factors between $\chi_1$ and $\chi_{-1}$:
\begin{equation}
M_{ab} = \int d^2\vec{x}_\perp \gamma_{23}\partial_a \chi_1\partial_b(\chi_{-1} e^{-2i\phi}).
\end{equation}
The emerging mass terms in (\ref{ModEffAxial}) are again the result of the breaking of $SO(3)_{L_{\rm int}} \times SO(3)_{L_{\rm ext}} \rightarrow SO(3)_L$.  Thus only a certain combination of the internal rotational moduli $\vec{\lambda}_\perp^+$ and the translational $\vec{\xi}_\perp$ moduli remain massless.  Additionally, only the quantized mode associated with this combination will be gapless.  Thus since the gapless mode associated with the axial $U(1)_A$ symmetry is not broken by these additional terms, we have a total of two gapless excitations.  The non-Abelian mode becomes quasi-gapless.  This is a familiar result similar to the case of superfluid $^4$He.

Although we will not consider the axial $U(1)_{A+J^{\rm int}_z}$ asymmetric case in detail here (see \cite{Volovik:1985a} for a detailed discussion), we will make a few brief remarks on the number and type of modes existing for both the $\gamma_{2,3} = 0$ and $\gamma_{2,3} \neq 0$ cases.  The most important change that is observed for this case is that the modes $\omega_z$ and $\delta$ are no longer equivalent and thus the single mode for the axial symmetric case will divide into two type B modes with linear dispersion for the axial asymmetric case.  The other modes from the axial symmetric case are the same for the asymmetric case and thus a total of four modes exist for the axially asymmetric vortex.  Again for the case that $\gamma_{2,3} \neq 0$ but small the non-Abelian mode from $\omega_{x,y}$ will become quasi-gapless.  Table 2 summarizes the number and type of modes emerging for the various core types.

\begin{table}
\caption{\label{Table2}We indicate the core type along with the type and number of quantized modes existing on the mass vortex with the corresponding core type.  In columns 2-4 we show whether the modulus exists as a quantized mode, and whether that mode has quadratic (type A) or linear (type B) dispersion.  A star indicates that the mode becomes quasigapless for $\gamma_{2,3} \neq 0$.} 
\begin{indented}
\item[]\begin{tabular}{@{}lllllcc}
\br
Core Type & $\vec{\xi}_{\perp}$ & $\omega_{x,y}$ & $\omega_z$ & $\delta$ & \#\ Modes $\gamma_{2,3} = 0$  & \#\ Modes $\gamma_{2,3} \neq 0$\\ 
\mr
$\vec{\chi}=0$, $f = f(r)$ & A & $\times$ & $\times$ & $\times$ & 1 & 1\\
$\vec{\chi}=0$, $f = f(x,y)$ & A & $\times$ & $\times$ & B & 2 & 2\\
$\vec{\chi}=\chi_z(r)\hat{z} \in \mathbb{R}^3$ & A  &  $\times$  & $\times$ & B & 2 & 2\\ 
$\vec{\chi}=\chi_z(r)\hat{z} \in \mathbb{C}^3$$^{\rm a}$ & A  &  B$^\star$  & $\times$ & B & 4 & 2\\ 
$U(1)_{A+J^{\rm int}_z}$ symmetric & A & A$^{\star}$ &  \multicolumn{2}{c}{${\rm B}  \sim {\rm B}$$^{\rm b}$} & 3 & 2\\
Broken $U(1)_{A+J^{\rm int}_z}$ & A & A$^{\star}$ & B & B & 4 & 3 \\
\br
\end{tabular}
\item[] $^{\rm a}$We assume the ground state of $\vec{\chi} \in \mathbb{R}^3$ along the $z$-axis with $\vec{h} \in \mathbb{C}^3$ excitations.
\item[] $^{\rm b}$ We have noted the equivalence of $\delta$ and $\omega_z$ modes for the $U(1)_{A+J^{\rm int}_z}$ core.
\end{indented}
\end{table}

\section{Conclusions}
In this work we have extended the analysis of previous analysis to include a description of gapless excitations of axially symmetric vortices in systems similar to the Ginzburg-Landau description of superfluid $^3$He in the B phase.  We have shown that in the case that $\gamma_{23} =  0$ the classical low energy theory includes the translational modulus fields presenting the well studied Kelvin excitations (see \cite{Thomson:1880, Sonin:1987zz, Simula:2008a, Fonda, Krusius:1984a, Pekola:1985a, Hakonen:1983a, Kobayashi:2013gba}).  In addition, the theory contains three additional moduli from the breaking of the non-Abelian $U(1)_A \times SO(3)_{J_{\rm int}}$ to the locked orbital-phase $U(1)_{A + J^{\rm int}_z}$ group.  These results have been developed using standard techniques of high energy physics to determine the classical low energy theory.

In addition, we have introduced the time dependent part of the Ginzburg-Landau free energy to facilitate the discussion of quantization of the gapless excitations.  In particular, we have discussed the number of gapless excitations emerging from the classical moduli determined from the broken symmetry generators.  In the present context the procedure of determining the gapless Goldstone modes is somewhat subtle due to the non-Lorentz invariance of the action \cite{Nielsen:1976, Watanabe:2012, Hidaka:2012ym}.  Specifically, we have shown that out of the five moduli emerging from the broken symmetry, only three appear as independent Goldstone modes after quantization.   Two of these are the type A Kelvin mode and the type B axial $U(1)_A$ mode, which have been well studied both experimentally and theoretically \cite{Thomson:1880, Sonin:1987zz, Simula:2008a, Fonda, Krusius:1984a, Pekola:1985a, Hakonen:1983a}.  The additional mode is the type A internal non-Abelian mode resulting from the breaking of $SO(3)_{J_{\rm int}}$ by the vortex line.

Additionally, upon introduction of small but non-zero $\gamma_{23}$ terms in the free energy (\ref{GLFE}) we observe the translational and rotational $SO(3)_{J_{\rm int}}$ modulus fields develop additional interactions.  This is due to the explicit breaking of the separate internal and external rotations $SO(3)_{L_{\rm int}} \times SO(3)_{L_{\rm ext}}$ to the rotational $SO(3)_{L}$.  Since external rotations of the vortex line about the $x$ and $y$ axes are equivalent to local translations we see that the translational moduli couple to the $SO(3)_{J_{\rm int}}$ moduli with coefficients  proportional to $\gamma_{23}$ in the free energy.  In addition, the new terms imply that only a particular combination of the translational and non-Abelian moduli remain massless.  This process is sometimes called the inverse Higgs mechanism in the high energy community \cite{Ivanov:1975, Clark:2003, Low:2001bw, Nitta:2013mj}.  In this case only the moduli from the broken Abelian symmetries remain strictly massless, while the other two non-Abelian modes are considered quasi-gapless.  Thus after quantization only the Kelvin and $U(1)_A$ mode remain strictly gapless.  The non-Abelian mode becomes quasi-gapless for small $\gamma_{2,3}$.

We wish to point out that although we have extended our analysis from the previous paper \cite{Peterson:2013zba}) to include the effects of quantization on our classically gapless modes, we have neglected to include the effects of higher order loop corrections to the effective potential.  In the context of condensed matter systems such as superfluid $^3$He these corrections will have little effect on the conclusions of the present work.  However, for an application of these corrections to the case of non-abelian strings see the analysis presented in \cite{Nitta:2013wca}.

We wish to conclude by pointing out the similarities between the present discussion of the excitations of mass vortices in condensed matter systems to the excitations of ANO  strings \cite{Abrikosov:1957, Nielsen:1973} in Yang-Mills theories.  The low energy theory describing the gapless excitations of axially symmetric vortices closely resembles the emergent $1+1$-dimensional O(3) sigma model of non-Abelian modes of ANO strings in Yang-Mills theories \cite{Monin:2013kza}.  In the present case however, the $U(1)_P$ phase symmetry is considered as a global symmetry in contrast to the case of Yang-Mills ANO strings where the phase symmetry is a $U(1)$ gauge symmetry.  Thus in our case the phase symmetry presents a modulus field on the low energy theory, which would not be the case for ANO strings in Yang-Mills theories.  This modulus $\delta$ leads to the $U(1)_A$ Bogoliubov mode with linear dispersion.

\ack
The authors would like to thank G. Volovik and M. Nitta for their comments and suggestions.  A. P. would like to thank Gianni Tallarita for interesting discussions.  M. S. is grateful to A. Kamenev for his comments.  This work was supported in part by DOE grant DE-FG02-94ER40823.

\end{document}